\documentclass[reprint,amsmath,amssymb,aps,fixfloat]{revtex4-2} 

\usepackage{graphicx}
\usepackage{dcolumn}
\usepackage{bm}

\newif\ifdraft

\ifdraft
\usepackage{svn}
\SVN $Id: Lee-Letter-2022.tex 405 2023-04-14 18:01:20Z shadwick $

\usepackage[color]{showkeys}

\definecolor{refkey}{rgb}{0.609,0.698,1}
\makeatletter
\def\SK@@ref#1>#2\SK@{%
  {\@inlabelfalse\leavevmode\vbox to\z@{%
	\vss
	\SK@refcolor
	\rlap{\raise0.85em%
	   \hbox{{\normalfont\fontsize{5.5pt}{0pt}\selectfont#2}}}}}}
\makeatother
\def\versionid{\footnote{version: \SVNId}}
\else
\let\versionid\relax
\fi

\allowdisplaybreaks

\def\Re{\mathop{\mathrm{Re}}\nolimits}
\def\Im{\mathop{\mathrm{Im}}\nolimits}
\def\sech{\mathop{\mathrm{sech}}\nolimits}
\def\W{\mathop{\mathrm{W}}\nolimits}
\def\Pn{\mathop{\mathrm{P}_n}\nolimits}
\def\Resn{\mathop{\mathrm{Res}_n}\nolimits}

\def\erfc{\mathop{\mathrm{erfc}}\nolimits}
\def\rhoi{\rho_{\hskip0.5pt\mathrm{ ion}}}

\def\Xip{\Xi^+}
\def\Xim{\Xi^-}
\def\Xipm{\Xi^\pm}
\def\Fp{F^+}
\def\Fm{F^-}
\def\Fpm{F^\pm}
\def\Phip{\Phi^+}
\def\Phipm{\Phi^\pm}
\def\Phim{\Phi^-}

\def\deltaE{\delta \mskip-1.5mu E}

\def\df{f^{(1)}}
\def\dE{E^{(1)}}

\def\textlabel#1{{\mathrm{#1}}}
\def\EvK{E_\textlabel{vK}}
\def\EvKC{\dE_\textlabel{vKC}}
\def\EJ{\dE_\textlabel{J}}
\def\DJ{D_\textlabel{J}}
\def\Eu{E_\textlabel u}
\def\Ed{E_\textlabel d}
\def\EL{E_\textlabel L}
\def\EW{E_\textlabel W}

\def\wp{\omega_p}
\def\lambdaD{\lambda_\mathrm{D}}
\def\vth{v_\mathrm{th}}

\def\bgroupm{\bgroup\color{magenta}}
\def\bgroupr{\bgroup\color{red}}
\let\egroupm\egroup

\def\EO{\mathcal{E}}
\def\EOt{\widetilde{\mathcal{E}}}

\begin{document}


\title{A Novel Method for Solving the Linearized 1D Vlasov--Poisson Equation}

\author{Frank M. Lee}
\email{flee3@unl.edu}
\author{B. A. Shadwick}
\email{shadwick@unl.edu}
\affiliation{University of Nebraska-Lincoln, Lincoln, NE 68588, United States}

\date{\today}

\begin{abstract}
We present a novel method for solving the linearized Vlasov--Poisson equation, based on analyticity properties of the equilibrium and initial condition through Cauchy-type integrals, that produces
algebraic expressions for the distribution and field, i.e., the solution is expressed without integrals.  Standard extant approaches involve deformations of the Bromwich contour that give erroneous
results for certain physically reasonable configurations or eigenfunction expansions that are misleading as to the temporal structure of the solution.  Our method is more transparent, lacks these
defects, and predicts previously unrecognized behavior.\versionid%
\end{abstract}

\maketitle

The 1D Vlasov--Poisson equation,
\begin{equation}
\frac{\partial f}{\partial t} + v \frac{\partial f}{\partial x} + \frac{q}{m} E \frac{\partial f }{\partial v} = 0, \label{vlasov}
\end{equation}
describes a plasma of particles with charge $q$ and mass $m$ as a one-particle distribution function, $f(x, v, t)$, ignoring correlations \cite{Krall:1973aa}.  The electric field $E(x,t)$ is given by
\begin{equation}
\frac{\partial E}{\partial x} = 4 \pi \left( q \int_{-\infty}^\infty f dv + \rhoi\right),
\end{equation}
where $\rhoi$ is the ion charge density.  We linearize about a spatially uniform equilibrium, $f_0(v)$, with a fixed, neutralizing ionic background, i.e., we take $f = n_0 \left[ f_0(v) +
\delta\!f(x,v,t) \right]$, where $\delta\!f$ is a small perturbation and $n_0$ is the equilibrium number density.  While we 
generically expect the linearization to only hold for a limited time, either due to
growth of the electric field in the case of an instability or due to phase space filamentation~\cite{Krall:1973aa} in the stable case resulting in large velocity gradients, we nonetheless consider the
linearized equation as a free-standing theory.  The resulting perturbed electric field, $\deltaE$, is given by
\begin{equation}
\frac{\partial }{\partial x} \deltaE = 4 \pi q \,n_0 \int_{-\infty}^\infty \delta\!f \, dv.\label{gl-linear}
\end{equation}
The linearized Vlasov equation,
\begin{equation}
\frac{\partial }{\partial t} \delta\!f + v \frac{\partial }{\partial x} \delta\!f + \frac{q }{m}\,f_0' \, \deltaE = 0, \label{vlasovlinear}
\end{equation}
has received considerable attention in the literature as an exact solution is possible.  Extant methods generally fall into two broad categories: the approach of van Kampen~\cite{vanKampen55} and
Case~\cite{Case59,Case78}; and that of Jackson~\cite{Jackson:1960aa} (extending Landau~\cite{Landau:1946aa}).  Here we introduce a new method that overcomes limitations in these approaches and we
resolve an apparent contradiction between the methods for the case of unstable equilibria.  Our solution, using well-established properties of Cauchy-type integrals~\cite{Gakhov}, only involves
Laurent series expansions and algebraic manipulations in the complex plane and has exact representations free of integral expressions, unlike the van Kampen--Case solution.  Furthermore, our method
can naturally yield a correct asymptotic approximation to the field and distribution, unlike the Jackson solution.
 
Case \cite{Case59,Case78} generalizes van Kampen's~\cite{vanKampen55} method by treating the linearized Vlasov equation as an eigenvalue problem in which zeros of the dielectric function are discrete
eigenvalues and van~Kampen's stationary waves correspond to real continuous eigenvalues, whose contribution is written as a continuum integral.  Since the van~Kampen contribution to the solution is
left as an opaque integral expression, this formulation obscures its temporal behavior.  Jackson \cite{Jackson:1960aa} expands on the approach of Landau~\cite{Landau:1946aa}, by shifting the Laplace
inversion contour and analytically continuing the integrand, writing the electric field for all time as the sum of residues.  Jackson's approach relies on the assumption of a ``reasonable'' initial
perturbation, which, upon careful analysis, we find to be overly restrictive.  Additionally, Jackson's method suggests that all solutions are sums of exponentials with arguments linear in time, which
is, as we will show below, not true for at least one class of initial conditions.  Jackson's method, widely cited by standard textbooks, fails to capture the entire time evolution by overlooking
functional properties in the complex plane and in general does not give a correct asymptotic approximation to the solution.

Taking $\df(k,v,t)$ and $\dE(k,t)$ to be the spatial Fourier transforms of $\delta\!f(x,v,t)$ and $\deltaE(x,t)$, respectively, \eqref{vlasovlinear} becomes
\begin{subequations}
	\label{linearvlasov}
	\begin{equation}
		\left(\frac{\partial}{\partial t} + ivk \right) \df(k,v,t) + \frac{q}{m}\,f_0'(v)\,\dE(k,t) = 0\,,\label{linf}
	\end{equation}
	where $\dE(k,t)$ is obtained from the Fourier transform of~\eqref{gl-linear}:
	\begin{equation}
		\dE(k,t) = \frac{ \EO}{ik\lambdaD}\int_{-\infty}^\infty \df(k,v,t)\,dv, \label{gauss}
	\end{equation}	
\end{subequations}
where $\EO= m\,\vth\,\wp/q$, $\wp = \sqrt{4 \pi q^2 n_0/m}$ is the plasma frequency, $\lambdaD = \vth/\wp$ is the Debye length, and $\vth$ is the thermal velocity of the equilibrium.  For
brevity we hereafter suppress the $k$ dependence.  When solving \eqref{linearvlasov}, two sectionally-analytic functions~\cite{Gakhov} arise naturally: the dielectric function,
\begin{equation}
    \epsilon(\omega) = 1 -\frac{\wp^2}{k^2} \!\!\int_{-\infty}^\infty\frac{f_0'(v)}{v - u}\,dv =
    \begin{cases}
        \Xip(u), &\Im \omega  > 0 \\
        \Xim(u), &\Im \omega < 0,
    \end{cases} 
	\label{epsilon}
\end{equation}
where $u=\omega/k$ and 
\begin{equation}
    \frac{1}{2 \pi i}\int_{-\infty}^\infty\frac{\df(v,0)}{v - u}\,dv = 
    \begin{cases}
		\Fp(u), &\Im u > 0 \\
		\Fm(u), &\Im u < 0,
    \end{cases} 
	\label{finitial}
\end{equation}
where $\df(v,0)$ is the initial perturbation.  The paired ``$+$'' and ``$-$'' functions on the right-hand sides of \eqref{epsilon} and \eqref{finitial} are sometimes referred to as ``Cauchy
splittings."  Using \eqref{gauss}, \eqref{epsilon}, and \eqref{finitial}, the solution of \eqref{linf} can be written as an inverse Laplace transform~\cite{Krall:1973aa},
\begin{equation}
	\df = \int_{\Gamma - i\infty}^{\Gamma + i\infty} \!\left[ \frac{\df(v,0)}{2 \pi i} + \frac{\wp^2 f_0'(v) }{k^2} \frac{F^+(is/k)}{\Xip(is/k)} \right]\!\frac{e^{st}\,ds}{s+ivk}\,, 
	\label{finverselaplace}
\end{equation}
where $\Gamma$ is chosen to place the contour to the right of all poles of the integrand, as required by the Laplace inversion theorem.  Likewise, the electric
field can be written as~\cite{Krall:1973aa}
\begin{equation}
	\dE = -\frac{\EO}{k^2\lambdaD^2} \! \int_{\Gamma - i\infty}^{\Gamma + i\infty}\frac{\Fp(is/k)}{\Xip(is/k)}\,e^{st}\, ds.\label{LaplaceE}
\end{equation}

For an unstable equilibrium, there appears to be a discrepancy between the methods of van Kampen--Case and Jackson.  A zero of $\epsilon$ in the upper half-plane implies a complex conjugate zero in
the lower half-plane, each giving a discrete mode.  The van Kampen--Case method includes contributions from the discrete modes in addition to those from modes associated with the continuous spectrum
(the so-called van Kampen modes).  The structure of \eqref{epsilon} and~\eqref{finitial} relate the amplitudes of each pair of discrete modes.  Thus, if $\epsilon$ has a single pair of simple zeros
$\omega_0$ and $\omega_0^*$, the electric field has the form:
\begin{eqnarray}
\EvKC = \frac{\EO}{ik\lambdaD} \left(A\,e^{-i\omega_0t} - A^* e^{-i\omega_0^*t}\right) + \EvK, \label{EvKC}
\end{eqnarray}
where
\begin{equation}
	\EvK = \frac{\EO}{i k\lambdaD}\int_{-\infty}^{\infty} du \left[\frac{\Fp{(u)}}{\Xip(u)} - \frac{\Fm(u)}{\Xim(u)} \right] e^{-ikut}
	\label{EvK}
\end{equation}
is the van Kampen mode contribution.  An unstable two-stream Maxwellian equilibrium,
\begin{equation}
	f_0 = \frac{1}{2\sqrt{2\pi}\,\sigma}\left[e^{-(v-v_0)^2/2\sigma^2} + e^{-(v+v_0)^2/2\sigma^2}\right],\label{ts-max}
\end{equation}
gives
\begin{multline}
\Xi^{\pm} = 1 + \frac{\wp^2}{k^2 \sigma^2}\bigg\{1 \pm i\,\sqrt{\frac{\pi}{2}} \bigg[\\
\frac{u + v_0}{2\,\sigma}\,\W{\left(\frac{u+v_0}{\pm \sqrt{2}\,\sigma}\right)} + \frac{u - v_0}{2\,\sigma}\,\W{\left(\frac{u - v_0}{\pm \sqrt{2}\,\sigma}\right)} \bigg] \bigg\},
\label{xipm2maxwellian}
\end{multline}
where
\begin{equation}
\W(z) = e^{-z^2}\erfc(-i\,z)
\end{equation}
is the Faddeeva function \cite{Oldham:2009aa}. The dielectric function corresponding to \eqref{xipm2maxwellian} has zeros $\omega_0 = \pm i g$, with $g>0$. Taking the initial
perturbation proportional to the equilibrium implies $A^* =  A$, giving
\begin{equation}
	\EvKC = \Eu + \Ed + \EvK\,, \label{EvKC2M}
\end{equation}
where
\begin{equation}
	\Eu = \frac{\EO}{ik\lambdaD}A\,e^{gt}\quad\mathrm{and}\quad \Ed = -\frac{\EO}{ik\lambdaD}A\,e^{-gt}. \label{EuEd}
\end{equation}
Jackson's method, however, only picks up contributions from the zeros of $\Xip$, the analytic continuation of $\epsilon$, as the Bromwich contour is shifted, deformed to encircle the poles, and
discarded in the limit $\Gamma\rightarrow -\infty$.  This implies an electric field of the form
\begin{equation}
\EJ = \Eu + \frac{\EO}{ik\lambdaD}\,\DJ, \label{EJ}
\end{equation}
where $\DJ$ is the contribution from the zeros of $\Xip$ in the lower half-plane.  The time evolutions implied by \eqref{EvKC2M} and \eqref{EJ} are distinctly different and it is thus important to
resolve this apparent discrepancy.

\begin{figure}[htb]
\includegraphics{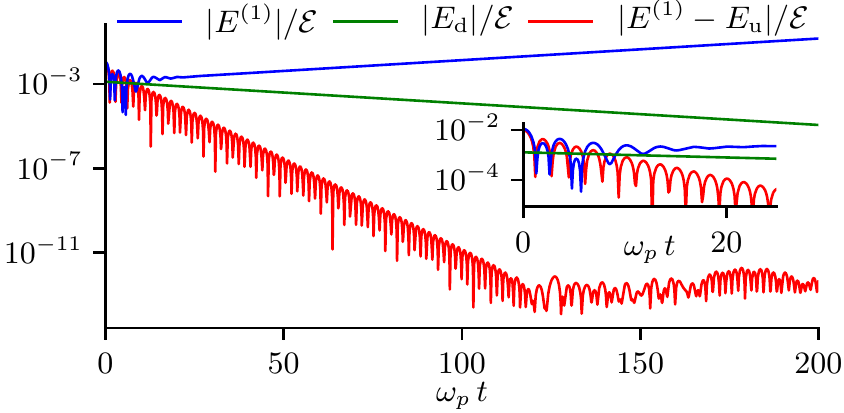}
\caption{\label{fig:manualdecay} The numerical solution of the linearized Vlasov--Poisson equation: the electric field $|\dE|/\EO$ (blue); the putative decaying mode contribution to the electric
field, $|\Ed|/\EO$ (green); and the electric field with the growing mode removed, $|\dE - \Eu|/\EO$ (red).}
\end{figure}

We solved the linearized Vlasov--Poisson equation numerically~\cite{Carrie:2022ws} on a periodic domain of length $L = 7\lambdaD$ with the equilibrium \eqref{ts-max}, $\sigma = \vth/2$, $v_0 =
\sqrt{3}\,\vth/2$, $\df(v, 0)= 0.01\,f_0$, and $k\,\lambdaD = 2\pi/7$.  Our results reveal that, contrary to Case, the decaying discrete mode is not present in the electric field; see
Fig.~\ref{fig:manualdecay}.  What remains after removing the unstable mode from the field (red curve) is clearly not dominated by the decaying discrete mode (green curve), but instead is the field
resulting from the zeros of $\Xip$ in the lower half-plane.

To gain more insight, consider a Lorentzian two-stream equilibrium for which the solution can be readily computed in closed form.  While, admittedly, this distribution is unphysical, the mathematics
of the linear theory is unaffected.  We take
\begin{equation}
	f_0 = \frac{\alpha}{2 \pi} \left[ \frac{1}{(v+v_0)^2 + \alpha^2} + \frac{1}{(v-v_0)^2 + \alpha^2} \right], \label{f0Lorentzians}
\end{equation}
for which
\begin{equation}
\Xi^\pm (u) = 1 - \frac{\wp^2}{k^2} \frac{ (u \pm i \alpha)^2 + v_0^2 }{ \left[ ( u \pm i \alpha )^2 - v_0^2 \right]^2 }. \label{epsilonLorentzians}
\end{equation}
Let $\beta = \wp\sqrt{\lambda - \Lambda^2 - 1/2}$, $\lambda^2 = 1/4 + 2\Lambda^2$, and $\Lambda = k\,v_0/\wp$.  If $v_0 > \alpha$, then for $k < \wp\sqrt{v_0^2 - \alpha^2 } / ( v_0^2 + \alpha^2 )$ we
have $\beta > k\,\alpha$ and $\epsilon$ has a root in the upper half-plane at $i \tilde{g} = i(\beta-k\,\alpha)$ and in the lower half-plane at $-i(\beta-k\,\alpha)$.  From~\eqref{epsilonLorentzians}
we see that the upper half-plane root corresponds to a zero of $\Xip$ while the lower half-plane root is a zero of $\Xim$.  Further, $\Xip$ has roots in the lower half-plane at frequencies $i h =
-i(\beta+k\,\alpha)$ and $\pm\delta -ik\,\alpha$, where $\delta = \wp\sqrt{\lambda + \Lambda^2 + 1/2}$; these are not zeros of $\epsilon$ but of its analytic continuation and, thus, \textit{are not
normal modes of the system}.

Since $\Xip$ has a finite number of zeros and the integrand in \eqref{LaplaceE} vanishes as $\Re s\rightarrow -\infty$, Jackson's method of taking the limit $\Gamma\rightarrow-\infty$ while encircling
the poles gives the correct solution in this case.  For this system, the integrals in the van Kampen--Case formalism can be readily computed, confirming the calculation based on~\eqref{LaplaceE}.  The
electric field is
\begin{equation}
\frac{\dE}{\EOt} = \mu\,\frac{\wp}{ik\alpha}\left(\frac{\zeta_+}2\,e^{\tilde{g}t} + \frac{\zeta_+}2\,e^{-ht} + \zeta_-\, e^{-\alpha t} \cos{\delta t}\right), \label{E2Lorentzian}
\end{equation}
where $\EOt = m\,\alpha\,\wp/q$, $\zeta_{\pm} = (\lambda \mp 2\Lambda^2 \mp 1/2 ) / (\lambda \mp 4\Lambda^2 \mp 1/2)$, and $\df(v,0) = \mu f_0(v)$.  (Note we choose the characteristic velocity scale
to be $\alpha$ instead of the undefined $v_{th}$.)  The first term in \eqref{E2Lorentzian} is the contribution from the root of $\epsilon$ in the upper half-plane, while the remaining terms are due to
zeros of $\Xip$ in the lower half-plane and represent Landau damping.  A closed-form expression for $\df(v,t)$ can be obtained but is unimportant here.  The form of \eqref{E2Lorentzian} is that of
\eqref{EJ} and not obviously \eqref{EvKC2M}; no term in \eqref{E2Lorentzian} is proportional to $e^{-\tilde{g}t}$, which would correspond to the root of $\epsilon$ in the lower half-plane.  The van
Kampen--Case prediction that the decaying discrete mode should be present in the solution is evidently incorrect.  In the calculation it becomes clear that $\EvK$ contains a term that
exactly cancels the decaying discrete mode contribution; as we show later, this cancellation is generic.

It is convenient to rotate the complex plane in \eqref{finverselaplace} with $u = is/k$ giving
\begin{equation}
	\df = \int_{i \gamma - \infty}^{i \gamma + \infty}\left[ \frac{\df(v,0)}{-2 \pi i} - \frac{\wp^2 f_0'(v) }{k^2} \frac{F^+(u)}{\Xip(u)} \right]\!\frac{e^{-ikut}}{u-v}\,du\,, 
	\label{fbromwich}
\end{equation}
where $\gamma = \Gamma/k$.  The contour is now horizontal and above the poles of the integrand.  The presence of the kernel $1/(u-v)$ in the integral allows us to apply well-established properties of
Cauchy-type integrals~\cite{Gakhov}.  We interpret~\eqref{fbromwich}, for $v$ real, as a ``$-$'' function resulting from Cauchy splitting about the Bromwich contour (recall that $\gamma>0$ and thus
the real $v$-axis lies below the Bromwich contour).  The strength of the method is that the Bromwich contour is left fixed and determining the unique ``$+$'' and ``$-$'' Cauchy splitting involves only
algebraic manipulation of the Cauchy density (the integrand excluding the kernel).  The distribution function becomes
\begin{equation}
    \df = \df(v,0) e^{-ikvt} - 2 \pi\,i\,\frac{\wp^2}{k^2}\,f_0'(v)\, \Phim(v, t), \label{f1}
\end{equation}
since the first term in \eqref{fbromwich} gives
\begin{eqnarray}
    \mskip-15mu\frac{1}{2 \pi i}\int_{i \gamma - \infty}^{i \gamma + \infty} \frac{e^{-ikut}}{u-z}\,du =
    \begin{cases}
        \ 0, &\Im z > \gamma \\
        \ -e^{-ikzt}, &\Im z < \gamma
    \end{cases} \label{fbromwich1}
\end{eqnarray}
and we have defined
\begin{eqnarray}
    \mskip-15mu\frac{1}{2 \pi i}\int_{i \gamma - \infty}^{i \gamma + \infty}\frac{\Fp(u)}{\Xip(u)}\,\frac{e^{-ikut}}{u-z}\,du  =
    \begin{cases}
        \Phip, &\Im z > \gamma \\
        \Phim, &\Im z < \gamma.
    \end{cases} \label{fbromwich2}
\end{eqnarray}
We find $\dE(t)$ from~\eqref{linf} using \eqref{f1}:
\begin{equation}
     \dE = 2\pi\,i\, \frac{\EO}{k^2\lambdaD}\left(\frac{\partial}{\partial t} + ikv\right)\Phim. \label{E}
\end{equation}
Thus, the key to computing the solution is determining $\Phim$, given $\Fp e^{-ikut}/\Xip$.  Whereas $F^{\pm}$ and $\Xi^{\pm}$ are Cauchy splittings about the real axis, $\Phi^{\pm}$ are
Cauchy splittings about the Bromwich contour.  Thus $\Fm$ or $\Xim$ are not used in evaluating $\Phim$; the problem is framed entirely by $\Fp$, $\Xip$, and $e^{-ikut}$
and their respective analyticity properties.

The splitting \eqref{fbromwich2} must respect the appropriate functional properties of $\Phipm$: they must be analytic and vanish as $\Im z\rightarrow\pm\infty$.  Note, no such conditions exist for
$\Im z\rightarrow\mp\infty$.  A Sokhotski relation~\cite{Gakhov} gives,
\begin{eqnarray}
\frac{\Fp(u)}{\Xip(u)} e^{-ikut} =\Phip(u, t) - \Phim(u, t). \label{fsokhotski}
\end{eqnarray}
The behavior of $\Fp$ below the real axis is the crucial fact that Jackson's construction ignores.  However, if we assume $\Fp e^{-ikut}/\Xip$ vanishes as $\Im u\rightarrow -\infty$ (as Jackson had
done), the only singularities are from the poles of $\Fp$ and the zeros of $\Xip$.  Then, $\Phi^{\pm}$ can be deduced by a separation of the poles from $\Fp/\Xip$:
\begin{equation}
	\Phim = -\frac{\Fp(z)}{\Xip(z)} e^{-ikzt} + \sum\limits_n \Pn\left[ \frac{\Fp(u)}{\Xip(u)} e^{-ikut} \right](z), \label{phimlorentzian}\\
\end{equation}
and 
\begin{equation}
	\Phip =  \sum\limits_n \Pn\left[ \frac{\Fp(u)}{\Xip(u)} e^{-ikut} \right](z),\label{phiplorentzian}
\end{equation}
where $\Pn[\varphi](z)$ denotes the principal part of the Laurent series of $\varphi$ about its $n$-th pole as a function of $z$.  Inserting $\Phim$ \eqref{phimlorentzian} into \eqref{f1} and
\eqref{E} gives
\begin{multline}
    \df = \df(v,0)\,e^{-ikvt} + 2\pi\,i\, \frac{\wp^2}{k^2}\,f_0'(v)\,\Biggl\{\frac{\Fp(v)}{\Xip(v)}\,e^{-ikvt} \\
    - \sum_n \Pn\left[\frac{\Fp(u)}{\Xip(u)}\,e^{-ikut}\right](v)\Biggl\} \label{florentzian}
\end{multline}
and
\begin{equation}
    \frac{\dE}{\EO} = -\frac{2\pi}{k\lambdaD}\sum_n\Resn\left[\frac{\Fp(u)}{\Xip(u)}\,e^{-ikut}\right], \label{Elorentzian}
\end{equation}
where $\Resn[\varphi]$ denotes the residue of $\varphi$ at its $n$-th pole.  Note, the sums may have an infinite number of terms.  Expressions \eqref{florentzian} and \eqref{Elorentzian} are the
conventional Jackson result obtained by shifting and deforming the contour around the poles, which is widely cited by standard textbooks.  However, this result is only correct provided $\Fp/\Xip$
diverges more slowly than $e^{ikut}$ as $\Im u\rightarrow -\infty$.

A Maxwellian initial condition gives an $\Fp$ that does not generally satisfy this condition and requires a modification to the Cauchy splitting \eqref{fsokhotski} and hence the solution will differ
from that from Jackson's formalism.  Consider
\begin{equation}
\df(v,0) = \frac{\mu}{\sqrt{2\pi}\,a}\,e^{-v^2/2a^2},
\end{equation}
which has essential singularities in both half-planes
and can be separated as
\begin{equation}
\df(u,0) = \frac{\mu}{\sqrt{8\pi}a} \left[\W\left(\frac{u}{\sqrt{2}\,a}\right) + \W\left(-\frac{u}{\sqrt{2}\,a}\right) \right]. \label{f10maxwellian}
\end{equation}
The first (second) term on the right-hand side of \eqref{f10maxwellian} is analytic in the upper (lower) half-plane and vanishes at infinity but has an essential singularity at infinity in the
opposite half-plane.  Thus,
\begin{equation}
\Fpm = \pm\frac{\mu}{\sqrt{8\pi}\,a}\W\left(\pm\frac{u}{\sqrt{2}\,a}\right)
\end{equation}
and separating the poles due to $\Xip$ and the singularities due to $\W$ and $e^{-ikut}$, which require shifting the argument, allows for a Cauchy splitting \eqref{fsokhotski}, giving
\begin{multline}
	\Phim = \frac{\mu}{\sqrt{8\pi}\,a} \bigg\{ \W\left(\frac{-z}{\sqrt{2}a} \right)  \!\frac{e^{ - ikzt}}{\Xip} -\W\left(\frac{ibt - z}{\sqrt{2}a}\right) \!\frac{e^{ - ct^2}}{\Xip} \\
	- \sum_n \Pn \!\bigg[ \W\!\left( \frac{-u}{\sqrt{2}a} \right) \!\frac{e^{ - ikut}}{\Xip}
	-\W\!\left(\frac{ibt - u}{\sqrt{2}a}\right) \!\frac{e^{ - ct^2 }}{\Xip} \bigg](z)\bigg\}, \label{Phimgaussian}
\end{multline}
where $b = -ka^2$ and $c = k^2a^2/2$.  When this is inserted into \eqref{E}, we find
\begin{multline}
	\frac{\dE}{\EO} = \frac{\mu}{i k\lambdaD} \bigg\{\sqrt{\frac\pi2}\,\frac ia\,\sum_n \Resn \bigg[ \W\left( \frac{-u}{\sqrt{2}a} \right) \frac{e^{ - ikut}}{\Xi^+} \\
	-\W\left(\frac{ibt - u}{\sqrt{2}a}\right) \frac{e^{ - ct^2}}{\Xip} \bigg]+ \Delta\,e^{-c t^2}  \bigg\}, \label{Egaussian}
\end{multline}
where $\Delta$ is a constant that depends on the equilibrium.  The more complicated behavior of the Maxwellian in the complex plane results in a fundamentally different Cauchy splitting and a time
behavior that is not simply a sum of complex exponentials with arguments linear in time, unlike what is expected from the Jackson method.  In this case, it is clear that as the Bromwich contour in
\eqref{LaplaceE} is shifted, $\Gamma\rightarrow-\infty$, the contour contribution to the integral does not vanish.

\begin{figure}[h]
\includegraphics{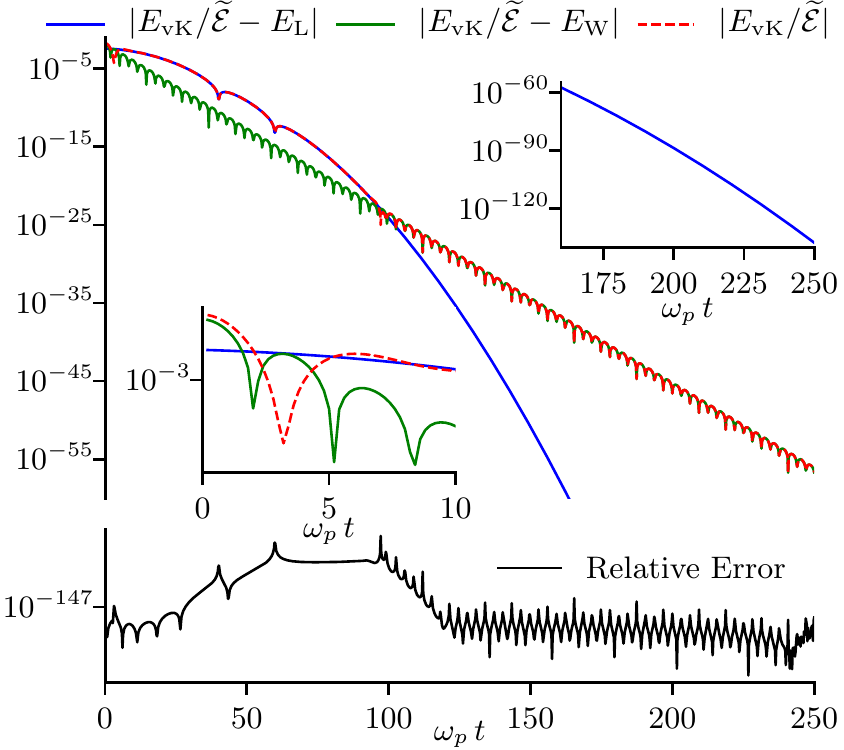}
\caption{\label{fig:comparison_vk} The electric field computed using van Kampen (dashed red) for a narrow Maxwellian initial perturbation and a Lorentzian equilibrium.  The relative error of fitting
the van Kampen solution to \eqref{fit}, shown in the lower panel, confirms the correctness of \eqref{EGausLorentz}.  See text for details.}
\end{figure}

To illustrate the consequences of \eqref{Egaussian} we take the equilibrium
\begin{equation}
f_0 = \frac{\alpha}{\pi}\frac{1}{v^2+\alpha^2},
\end{equation}
for which
\begin{equation}
\Xipm = 1 - \frac{\wp^2}{k^2} \frac{1}{(u \pm i\,\alpha)^{2}}
\end{equation}
and $\Delta = 1$.  The field is
\begin{multline}
\frac{\dE}{\EOt} = \mu\,\frac{\wp}{i k \alpha}\left(\eta\,\left|\W\left(\frac{ik\alpha - \wp}{\sqrt{2}ak}\right)\right|\, e^{-k\alpha t} \cos\left(\omega_p t + \theta\right)\right.\\[2pt]
\left.{}+e^{-c t^2 } \left\{ 1 + \eta\, \Im\left[\W \left(\frac{i bk t +  ik\alpha - \wp}{\sqrt{2}ak}\right)\right]\right\}
 \right)\label{EGausLorentz}
\end{multline}
where $\eta = \sqrt{2\pi}\,\wp/ak$ and $\theta = \arg\,\W[(ik\alpha - \wp)/\sqrt{2}ak]$.  The exact solution, with $\mu = 0.01$, $a/\alpha = 0.2$, and $k\,\alpha/\wp=1/2$, is compared with van Kampen's
solution in Fig.~\ref{fig:comparison_vk}.  We computed the van Kampen solution
by numerically evaluating~\eqref{EvK} using the arbitrary-precision Python library \texttt{mpmath}~\cite{Johansson:2013aa} carrying 200 decimal digits.  We compare \eqref{EvK} with the analytical
form \eqref{EGausLorentz} by fitting the electric field to $\EL + \EW$ where
\begin{subequations}
	\label{fit}
	\begin{equation}
		\EL = l_1e^{l_2 t} \cos(l_3\,t + l_4)	
	\end{equation}
and
	\begin{equation}
		\EW = w_1\,e^{-w_2 t^2 }[ 1 + w_3\,\Im\W(i w_4 t + iw_5 + w_6)]\,.
	\end{equation}
\end{subequations}
Doing so requires surprisingly high precision since the terms involving 
$\W$ are slowly varying functions of their parameters and the Gaussian-like behavior persists for some time before the
exponential decay (Landau damping) associated with the single pair of roots of $\Xip$ becomes dominant.  The red dashed line in the upper panel of Fig.~\ref{fig:comparison_vk} shows $\EvK$.  The blue
line shows the electric field with $\EL$ removed while the green line shows the electric field with $\EW$ removed.  The parameters extracted by the fitting reproduced the analytical values to
approximately 150 digits.  The fit residual, shown as the relative error in the lower panel of Fig.~\ref{fig:comparison_vk}, has little structure, confirming the fitted function well-reproduces the
analytical form.  It is often noted that Landau damping represents the ``long-time'' behavior of the system.  Here the field amplitude must fall by some twenty orders of magnitude before the system
becomes dominated by this long-time behavior; Landau damping is almost invisible for roughly the first sixteen plasma periods.  This is entirely due to contributions from the initial condition, which
is in no way physically unreasonable.

\begin{figure}[h]
\includegraphics{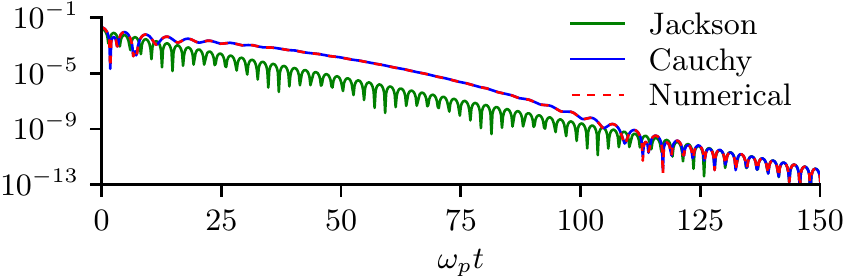}
\caption{\label{fig:gaussiangaussian} The electric field~\eqref{Egaussian} with~\eqref{xi+Max} (blue) for a Maxwellian initial perturbation narrower than the Maxwellian equilibrium.  Clearly,
Jackson's method \eqref{Elorentzian} (green) fails to capture a critical feature of the electric field that is present in the numerical solution (dashed red) and reproduced by the expressions obtained
via Cauchy splitting (blue).  See text for details.}
\end{figure}

It must be emphasized that the existence of non-exponential terms, \textit{i.e.}, those in additional to Jackson's result, does not depend solely on the form of either the initial condition or the
equilibrium, but whether $\Fp/\Xip$ diverges more quickly than $e^{ikut}$ as $\Im u\rightarrow -\infty$.  Thus the same initial condition as above and the equilibrium
\begin{equation}
f_0 = \frac{1}{\sqrt{2\pi}\,\vth} e^{-v^2/2\vth^2}
\end{equation}
can show qualitatively the same behavior
provided $a<\vth$.  For this equilibrium
\begin{equation}
	\Xip = 1 + \frac{\wp^2}{\vth^2 k^2} \left[ 1 + i\,\sqrt{\frac{\pi}{2}}\,\frac{u}{\vth}\,\W \left( \frac{u}{\sqrt{2}\,\vth} \right)\right], \label{xi+Max}
\end{equation}
and $\Delta = 3/4$.  Since $\Xip$ has an infinite number of zeros, it is not possible to produce a closed-form expression for the field.  The field, \eqref{Egaussian} with \eqref{xi+Max}, for
$a = \vth/10$ and evaluated for the first 1000 zeros of $\Xip$ is shown as the blue curve in Fig.~\ref{fig:gaussiangaussian}.  The green curve shows Jackson's solution evaluated with the same zeros.
The dashed red curve is a numerical solution of the linearized Vlasov--Poisson equation for this equilibrium and initial condition on a periodic domain of length $L = 4\pi\lambdaD$ with $k\lambdaD =
1/2$ and $\mu=0.01$.

\begin{figure}[h]
\includegraphics{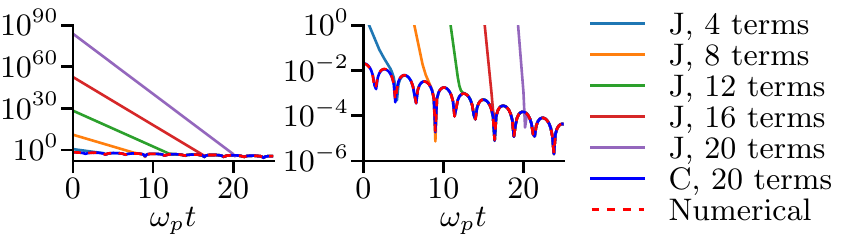}
\caption{\label{fig:gaussiansech} The electric field~\eqref{Egaussian} with~\eqref{xi+sech} (blue) and Jackson's solution \eqref{Elorentzian} truncated to various numbers of terms, left.  The right
panel shows an expanded scale.  Our solution by Cauchy-splitting (blue), taking only twenty terms, agrees very well with the numerical solution of the linearized Vlasov--Poisson equation (dashed
red).}
\end{figure}

Further, Jackson's method of shifting $\Gamma\rightarrow -\infty$ in \eqref{LaplaceE} while analytically continuing the integrand in general does not necessarily give a reliable asymptotic
approximation of the solution.  Consider the above initial condition and the equilibrium
\begin{equation}
f_0 = \frac{1}{2\,\vth}\sech\left(\frac{\pi\,v}{2\,\vth}\right)
\end{equation}
for which
\begin{equation}
\Xip = 1 + \frac{\wp^2}{8\,k^2\vth^2}\left[\psi^{(1)}\!\!\left(\frac{\vth - iu}{4\,\vth}\right) - \psi^{(1)}\!\!\left(\frac{3\vth - iu}{4\,\vth}\right)\right],\label{xi+sech}
\end{equation}
where $\psi^{(1)}$ is the trigamma function~\cite{Oldham:2009aa} and $\Delta = 1$.  Here Jackson's formulation fails since the solution diverges as more terms are added, as shown in
Fig.~\ref{fig:gaussiansech}.  The dashed red curve is a numerical solution of the linearized Vlasov--Poisson equation for this equilibrium and initial condition on a periodic domain of length $L =
4\pi\lambdaD$ with $k\lambdaD = 1/2$, $a = \vth$, and $\mu=0.01$.  The blue curve is~\eqref{Egaussian} with \eqref{xi+sech} using the first 20 zeros of $\Xip$.  This failure of Jackson's formulation
arises because the zeros of $\Xip$ progress down along the imaginary axis, which when evaluated in the numerator $\Fp$ in \eqref{Elorentzian} causes the amplitude to become more and more divergent.
In contrast, the numerator in~\eqref{Egaussian} from Cauchy splitting vanishes at infinity in the lower half-plane, resulting in a proper asymptotic form.  Thus it is crucial to examine the behavior
of the entire Cauchy density, not just $\Xip$ as is commonly done.

\begin{figure}[bth]
\includegraphics{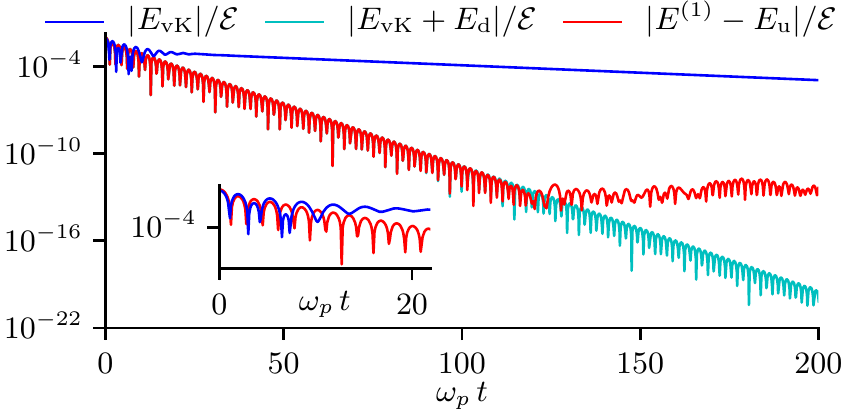}
\caption{\label{fig:vankampen} The contribution to the field from the van Kampen modes, $|\EvK|/\EO$ (blue), for the system shown in Fig.~\ref{fig:manualdecay}.  The field is dominated by
an exponential decay identical (but with opposite amplitude) to that of the Case decaying discrete mode, $\Ed$.  Adding $\Ed$ to the van Kampen contribution, $|\EvK+\Ed|/\EO$ (cyan), results in a
cancellation, giving the same evolution as when the growing mode is subtracted from the field from the numerical solution, $|\dE-\Eu|/\EO$ (red).}
\end{figure}

For an equilibrium with a finite collection of unstable modes of any order, the van Kampen--Case solution~\cite{vanKampen55,Case78,Case59}, in our notation, becomes
%
\begin{multline}
\df = -2 \pi i \frac{\wp^2}{k^2}f'_0 \Bigg\{ \sum_j \text{P}_j \Bigg[ \frac{\Fp}{\Xip} e^{-ikut} \Bigg](v) \\
+\sum_j \text{P}_j \Bigg[\frac{\Fm}{\Xim} e^{-ikut} \Bigg](v) \Bigg\} + 
\int_{-\infty}^{\infty} du \left[\frac{\Fp{(u)}}{\Xip(u)} - \frac{\Fm(u)}{\Xim(u)} \right] \\ 
\times\Bigg[ \frac{\wp^2}{k^2}\,\text{P}\,\frac{f'_0(v)}{v-u} + \frac{\Xip(u) + \Xim(u)}{2}\,\delta(v-u) \Bigg] e^{-ikut},
\label{CvKunstable}
\end{multline}
where the sums are over the zeros of $\Xip$ in the upper half-pane and $\Xim$ in the lower half-plane and $\text{P}$ denotes the principal value.  Using properties of the Cauchy splittings we have
\begin{multline}
\lim_{\Im z \rightarrow 0^+} \frac{1}{2\pi i} \int_{- \infty}^{\infty}\frac{\Fm(u)}{\Xim(u)}\,\frac{e^{-ikut}}{u-z}\,du  =\\ 
\sum_j \text{P}_j\!\left[\frac{\Fm(u)}{\Xim(u)} e^{-ikut}\right]\!(v) \label{limp}
\end{multline}
and
\begin{multline}
\lim_{\Im z \rightarrow 0^-} \frac{1}{2\pi i} \int_{- \infty}^{\infty}\frac{\Fm(u)}{\Xim(u)}\,\frac{e^{-ikut}}{u-z}\,du  = - \frac{\Fm(v)}{\Xim(v)}\,e^{-ikvt} \\[4pt]
+ \sum_j \text{P}_j\!\left[\frac{\Fm(u)}{\Xim(u)} e^{-ikut}\right]\!(v), \label{limm}
\end{multline}
which combined with Sokhotski's relations~\cite{Gakhov},
\begin{multline}
\lim_{\Im z \rightarrow 0^{\pm}} \int_{- \infty}^{\infty}\frac{\Fm(u)}{\Xim(u)}\,\frac{e^{-ikut}}{u-z}\,du  = \pm i \pi \frac{\Fm(v)}{\Xim(v)}\,e^{-ikvt} \\[4pt]
+ \mathrm{P}\!\! \int_{-\infty}^\infty\frac{\Fm(u)}{\Xim(u)}\,\frac{e^{-ikut} }{u - v}\, du, \label{limpm}
\end{multline}
allow us to write
\begin{multline}
\mathrm{P}\!\! \int_{-\infty}^\infty\frac{\Fm(u)}{\Xim(u)}\,\frac{e^{-ikut} }{v - u}\, du = i\pi\frac{\Fm(v)}{\Xim(v)} e^{-ikvt} \\[4pt]
- 2\pi\,i \sum_j \text{P}_j\!\left[\frac{\Fm(u)}{\Xim(u)} e^{-ikut}\right]\!(v). \label{pvm}
\end{multline}
When inserted into \eqref{CvKunstable}, the terms in the sum over $j$ in \eqref{pvm} exactly cancel the
decaying discrete mode contributions to the distribution function, giving
\begin{multline}
\df= -2 \pi i \frac{\wp^2}{k^2}f'_0 \sum_j \text{P}_j \Bigg[\frac{\Fp}{\Xip} e^{-ikut} \Bigg](v)\\
-i\pi\,\frac{\wp^2}{k^2}\,f'_0\,\frac{\Fm}{\Xim}\,e^{-ikvt} 
+ \int_{-\infty}^{\infty}\!\!du \Bigg\{\!\frac{\Fp{(u)}}{\Xip(u)} \frac{\wp^2}{k^2}\,\text{P}\,\frac{f'_0(v)}{v-u} \\
+\left[ \frac{\Fp{(u)}}{\Xip(u)} - \frac{\Fm(u)}{\Xim(u)} \right]\!\!\frac{\Xip(u) + \Xim(u)}{2}\,\delta(v-u)\!\Bigg\} e^{-ikut}. \label{cancelled}
\end{multline}
It is thus apparent that Case's counting of discrete modes~\cite{Case59,Case78} is not correct.  The decaying discrete modes, while necessarily present whenever there are growing discrete modes as
part of the eigenvalue problem, always cancel with a contribution from the van Kampen modes, and thus \textit{never contribute to the time evolution} of the distribution function or electric field.
This result is illustrated in Fig.~\ref{fig:vankampen}.

Elskens and Escande~\cite{Elskens:2003aa} developed a finite-degree-of-freedom particle approximation to Vlasov dynamics in the context of quasi-linear theory.  They argue that an unstable/stable
eigenmode pair is accompanied by Landau damping in such a way that the decaying eigenmode contribution to the electric field is cancelled by Landau damping.  In distinction, while we find that the
continuum does exactly cancel the decaying discrete mode, the rate of the discrete mode is significantly different than the decay rate from the Landau pole, and the same is true of the
phase velocities.  Our result is a rigorous property of the solution of the linearized system and not the consequence of an approximate argument.

We have developed a method for solving the linearized Vlasov--Poisson system, based on analyticity properties of the equilibrium and initial condition by exploiting properties of Cauchy-type
integrals.  Our method produces algebraic expressions for the distribution and field, i.e., the solution is expressed without integrals; this interpretation of \eqref{fbromwich2} can be generalized to
a wide class of inversion integrals.  When the equilibrium results in $\Xip$ having an infinite number of zeros, our method can yield a reliable asymptotic approximation to the field and distribution
function.  We showed that for ``reasonable'' initial conditions and stable equilibria, the solution may exhibit time dependence more complicated than simple exponential decay and oscillation as
expected from a naive evaluation of the inverse Laplace transform.  Our analysis explained an apparent discrepancy in unstable systems between the Jackson and van Kampen--Case solutions.  We showed
that the contribution of decaying discrete modes to the field and distribution is always canceled by a corresponding contribution from the van Kampen modes.  Case's formulation in terms of orthogonal
eigenfunctions overlooked the fact that the van Kampen modes and the decaying discrete mode eigenfunctions are not independent.  As a result, Case's ``mode counting'' is incorrect.  Not all of the
normal modes arising from solutions of $\epsilon=0$ appear in the solution; only the unstable discrete modes survive.

\begin{acknowledgments}
This work was supported by the U.\ S.~DoE under contract No.\ DE-SC0018363 and by the NSF under contract No.\ PHY-2108788.
\end{acknowledgments}
F.M.L. and B.A.S. contributed equally to this work.

\bibliography{refs}

\end{document}